**Research Article**

# Examining the effects of music on cognitive skills of children in early childhood with the Pythagorean fuzzy set approach

Murat KIRIŞCI[1], Nihat TOPAÇ[2,*], Musa BARDAK[3]

[1]*Department of Biostatistics and Medical Informatics, Istanbul University-Cerrahpasa, Istanbul, 34098, Türkiye*
[2]*Department of Early Childhood Education, Istanbul University-Cerrahpasa, Istanbul, 34098, Türkiye*
[3]*Department of Early Childhood Education, Istanbul Sabahattin Zaim University, Istanbul, 34303, Türkiye*

**ABSTRACT**

There are many genetic and environmental factors that affect cognitive development. Music education can also be considered as one of the environmental factors. Some researchers emphasize that Music is an action that requires meta-cognitive functions such as mathematics and chess, and supports spatial intelligence. The effect of music on cognitive development in early childhood was examined with the Pythagorean Fuzzy Sets(PFS) method defined by Yager. In this study, PFS was created from experts' opinions and an algorithm was given according to PFS. The results of the algorithm supported the data of the experts on the development of spatial-temporal skills of music education given in early childhood. In the algorithm, the ranking has been done with the Expectation Score Function. The rankings obtained from the algorithm overlap with the experts' rankings.

**Cite this article as:** Kırışcı M, Topaç N, Bardak M. Examining the effects of music on cognitive skills of children in early childhood with the Pythagorean fuzzy set approach. Sigma J Eng Nat Sci 2023;41(4):742–749.

## INTRODUCTION

**Pythagorean Fuzzy Environment**

Uncertainty is a crucial concept for decision-making problems. It is not easy to make precise decisions in life since each information contains vagueness, uncertainty, imprecision. Fuzzy Set(FS) Theory, Zadeh's [35] pioneering work, proposed a membership function to solve problems such as vagueness, uncertainty, imprecision, and this function took value in the range of [0,1]. In [10,11], analyzes were made using a Fuzzy matrix. FS Theory had solved many problems in practice, but there was no membership function in real life, which only includes acceptances. Rejection is as important as acceptance in real life. Atanassov [2] clarified this problem and posed the Intuitionistic Fuzzy Set (IFS) Theory using the membership function as well as the non-membership function. In IFS, the sum of membership and non-membership grades is 1. This condition is also a limitation for solutions of vagueness, uncertainty, imprecision. Yager [34] has presented a solution to this situation and suggested Pythagorean Fuzzy Sets (PFS). PFS is more comprehensive than IFS because it uses the condition that the sum of the squares of membership and non-membership grades is equal to or less than 1. Despite all the possible solutions, these theories have limitations. How to set the membership function in each particular object and the deficiencies in considering the parametrization tool can be given as examples of these limitations. These limitations handicap decision-makers from making a correct decision during the analysis.

A new method, called Soft Set (SS), was proposed by Molodstov [20], in which the preferences for each alternative were given in distinct parameters, and thus a solution was found for the limitations expressed. Immediately after the occurrence of SS theory, Fuzzy Soft Sets [16] and Intuitionistic Fuzzy Soft Set (IFSS) [17] were defined and their various properties were studied [18, 19]. Pythagorean Fuzzy Soft Set (PFSS) is defined by Peng et al. [24]. Soft sets have found a place in the literature with their various applications [7, 13, 14, 30]. PFSS is a natural generalization of IFSS and is a parameterized family of PFSs.

It is possible that people may hesitate during decision-making. In order to avoid human hesitations from adversely affecting the decision-making process, hesitation value is also taken into account in PFS, just like IFS. Thus, experts may have hesitations about membership grades. If the expert participating in the decision process is only one person, this expert's error or bias will affect the process negatively. However, hesitation is subjective and the expert's hesitation can be directed by her/his own perceptions. In this case, enriching the decision process, making the evaluation with alternative decisions more meaningful, combining the subjective evaluations of more than one expert instead of the subjective evaluation of a single expert will provide a healthier decision-making process. With this in mind, Agarwal et al. [1] defined Generalized Intuitionistic Fuzzy Soft Sets(GIFSS). Feng et. al. [7] identified some problems and difficulties in the definition of GIFSS and operations related to GIFSS in the manuscript of [1]. Kirisci [15] defined Generalized Pythagorean Fuzzy Soft Sets(GPFSS), considering the fixes in [7].

GPFSS ensures the frame for evaluating the reliability of the info in the PFSS so as to compensate for any distortion in the info given. The most important benefit of incorporation of the generalized parameter into the analysis is to decrease the likelihood of errors induced by the imprecise info by taking the chairperson's view on the same. For example, a patient may give wrong information to an expert about her/his symptoms. If the expert does not notice this wrong information, errors in diagnosis and treatment will occur. In this case, an experienced expert can measure the reliability of the information given by the patient with a generalization parameter. So, there is a requirement for a generalization parameter, demonstrating an expert's level of confidence in the reliability of the info, respectably making the approach quite close to real-world cases. This assists in extracting the singular bias from the input data and gets more credibility to the final decision. GPFSS has a generalization parameter to states the uncertainties. Nevertheless, GPFSS has some limitations as it is hard for a single senior/junior expert (or moderator) to serve an appropriate generalization parameter with fuzzier info according to his/her own knowledge.

IFS is characterized by a membership degree and a non-membership degree and therefore can indicate the fuzzy character of data in more detail comprehensively. The prominent characteristic of IFS is that it assigns to each element a membership degree and a non-membership degree with their sum equal to or less than 1. However, in some practical DM processes, the sum of the membership degree and the non-membership degree to which an alternative satisfying a criterion provided by a decision-maker may be bigger than 1, but their square sum is equal to or less than 1. Therefore, Yager [32] proposed PFS characterized by a membership degree and a non-membership degree, which satisfies the condition that the square sum of its membership degree and non-membership degree is less than or equal to 1. Here, we observe that intuitionistic membership grades are all points under the line $x + y \leq 1$ and the Pythagorean membership grades are all points with $x^2 + y^2 \leq 1$.

**Cognitive Development**

The period in which the development of the individual is the fastest as holistic is the early childhood period. In this period, the child is in the process of development in terms of social, emotional, language, psycho-motor, self-care, and cognitive aspects. In this process, the child is affected by various environmental conditions with formal or informal experiences. In this context, Vygotsky [31] emphasizes that cognitive development is significantly affected by the child's environment in terms of socio-culture. However, it has been determined by many studies that cognitive development has made great progress in early childhood. There are many factors that affect this progress. It can be said that one of these factors is music. Cooper [5] stated that studies on the cognitive benefits of music education are arousing a global scale. Cooper [5] stated that the studies on the cognitive benefits of music education are arousing curiosity on a global scale.

Cognitive development can be defined as the process of learning, practicing, and controlling cognitive skills that an individual can do using his mind. Oakley [23] cited cognitive skills as all processes related to learning, organizing, using, and developing knowledge. Executive skills such as being organizational, impulse control, planning, and working memory represent the basic cognitive functions of the person responsible for the ability to adapt to different (differentiated) environments [21]. Solso [29] described the cognitive field as a science in which brain functions such as perception, attention, memory, and thinking, language, problem-solving, and reasoning are examined. In addition to those in these definitions, it can be stated that self-regulation skills that include reasoning, problem-solving, and decision-making related to one's emotions, impulses, and thoughts are within the scope of cognitive development. Among the cognitive skills mentioned in the definitions, attention, memory, thinking and self-regulation skills were evaluated within the scope of this study.

It is known that it can benefit from different disciplines in order to support cognitive development. Katarzyna and Brenda [10] stated that the relationship between music and cognitive development has been studied by various researchers, including neuroscientists, psychologists, educational experts, and musicians. In addition, alternative methods and techniques are also used to support cognitive development outside of different disciplines. It can be said that some of these methods and techniques are methods such as play, drama, and music. In addition, different techniques such as finger game, rond, creative dance, and musicals have been created by combining these methods [3]. Perhaps the most important factor that supports cognitive development in music education is teacher qualifications. In addition to the knowledge of the needs of music education and the child's cognitive development, teachers should be well educated especially in the use of musical instruments and materials [30].

**Music Education in Early Childhood**

It is known that it can benefit from different disciplines in order to support cognitive development. There are many factors that affect this progress. It can be said that one of these factors is music. Cooper [5] stated that studies on the cognitive benefits of music education are arousing a global scale. Katarzyna and Brenda [12] stated that the relationship between music and cognitive development has been studied by various researchers, including neuroscientists, psychologists, educational experts, and musicians.

In addition, alternative methods and techniques are also used to support cognitive development outside of different disciplines. It can be said that some of these methods and techniques are methods such as play, drama, and music. In addition, different techniques such as finger games, rond, creative dance, and musicals have been created by combining these methods [3]. Perhaps the most important factor that supports cognitive development in music education is teacher qualifications. In addition to the knowledge of the needs of music education and the child's cognitive development, teachers should be well educated especially in the use of musical instruments and materials [26].

Since early times, many thinkers and educators have revealed the importance of music. For example, Comenius stated that the child was exposed to music from birth. He stated that in the next process, the child should be able to sing nursery rhymes and songs and discover sounds and instruments ([8]. In a study conducted on 5-year-old Japanese children, it was determined that music increases children's cognitive performance [26]. In another study, it was revealed that musical education given to children improves their spatial and temporal thinking skills [4] As we approach today, an increasing number of studies reveal that music education, starting from early childhood, significantly supports academic and social success [22].

In the early childhood period, many activities can be done for children within the scope of music education. Some of these are activities such as listening to and distinguishing sounds and music, rhythm exercises, breathing and voice exercises, singing, playing an instrument, creative movement and dance, movement with music, creating musical stories. It is revealed in different studies that these activities contribute directly and indirectly to all areas of development, especially cognitive development.

**Some Basic Concepts**

Some knowledge that will be used throughout the article will be given. Let $E, N$ be an initial universe and parameter sets, respectively.

**Definition 1** [20] For $A \subseteq N$ consider a set-valued mapping $I: A \rightarrow v(U)$, where the power set of $E$ is should by $v(U)$. Therefore, a pair $T = (I, A)$ is called a soft set(SS) on $E$.

**Definition 2** For $a \in E$, the set $B = \{(a, d_B(a), y_B(a)): a \in E\}$ on $E$ is called Pythagorean fuzzy set(PFS), where $dB: E \rightarrow [0,1]$ and $y_B: E \rightarrow [0,1]$ together the situation that $0 \leq d_B^2 + y_B^2 \leq 1$ [26, 27, 28]. The degree of indeterminacy $h_B = \sqrt{1 - [d_B(a)]^2 - [y(a)]^2}$.

**Definition 3** For $B \subseteq N$, choose $\tilde{F}: B \rightarrow v(E)$, where the set of all PFSs over $E$ is indicated by $v(E)$. Then, a pair $\tilde{T}_B := (\tilde{I}, B)$ is called Pythagorean Fuzzy Soft Set (PFSS) on $E$ [24].

Take $k = \{(r, s): r^2 + s^2 > 1, r, s \in [0,1]\}$. Let $(K, \leq_K)$ be a complete lattice. The corresponding partial order $\leq_K$ is defined by

$(r, s) \leq_K (t, u) \Leftrightarrow r \leq t$ and $s \geq u$.

For all $(r, s), (t, u) \in K$. The Pythagorean fuzzy value(PFV) is denoted by an ordered pair $(r, s) \in K$ [15].

**Definition 4** [15] Take the PFS $C$ over $N$. Let the PFS $f: C \rightarrow K$. If the set

$T_f = \{(a, d_C(a), y_C(a), f_c(a)): a \in P, f_C(a) \in K, d_C(a), y_C(a) \in [0,1]\}$

is PFSS on $E$, then the $\tilde{T}_f$ is called a Pythagorean fuzzy parameterized Pythagorean fuzzy soft set ($\Omega$-soft set). The elements of parameter $f_C$ are indicated by $(d_f, y_f)$. Here, $\Omega(E)$ shows the set of all $\Omega$-soft set on $E$.

**Example 1** Let's choose four experts who work in Early School Education. The experts are studying research on the effect of music on the cognitive development of the early childhood period. Let's take the set $P = \{p_1, p_2, p_3, p_4\}$ as the set of experts. In this research, they examine the following situations:

($S_1$) Music education improves children's attention capacity.

($S_2$) Music education improves children's vocabulary knowledge.

($S_3$) Music education improves children's reasoning and logical thinking skills.

($S_4$) Music education improves children's self-regulation skills.

for the set $S = \{s_1, s_2, s_3, s_4\}$. Then,

$F(s_1) = \{(p_1, 0.7, 0.7), (p_2, 0.5, 0.6), (p_3, 0.9, 0.4), (p_4, 0.7, 0.5)\}$,

$F(s_2) = \{(p_1, 0.6, 0.6), (p_2, 0.4, 0.9), (p_3, 0.8, 0.4), (p_4, 0.6, 0.5)\}$,

$F(s_3) = \{(p_1, 0.8, 0.2), (p_2, 0.8, 0.6), (p_3, 0.6, 0.7), (p_4, 0.5, 0.8)\}$,

$F(s_4) = \{(p_1, 0.4, 0.7), (p_2, 0.5, 0.6), (p_3, 0.7, 0.4), (p_4, 0.8, 0.3)\}$.

All this information can be represented in terms of the $F_P$ as table in Table 1. The values given in the table are arranged according to the opinions of the experts given in the literature.

**Table 1.** $F_p$

| P\S | s1 | s2 | s3 | s4 |
| --- | --- | --- | --- | --- |
| p1 | (0.7, 0.7) | (0.6, 0.6) | (0.8, 0.2) | (0.4, 0.7) |
| p2 | (0.5, 0.6) | (0.4, 0.9) | (0.8, 0.6) | (0.5, 0.6) |
| p3 | (0.9, 0.4) | (0.8, 0.4) | (0.6, 0.7) | (0.7, 0.4) |
| p4 | (0.7, 0.5) | (0.6, 0.5) | (0.5, 0.8) | (0.8, 0.3) |

**Definition 5** For $C, D \subseteq N$, choose two PFSS $\tilde{T}_C$ and $\tilde{T}_D$ such that $\tilde{T}_C \subseteq \tilde{T}_D$ and two Ω-soft sets $\tilde{T}_f$ and $\tilde{T}_g$. If $\tilde{T}_f$ is a Ω-soft subset of $\tilde{T}_g$ for all $u \in N$, then the following conditions are hold:
  i.   $\tilde{T}_C \subseteq \tilde{T}_D$ and $f_C \leq_K g_D$,
  ii.  $d_f(u) \leq d_G(u)$, $y_f(u) \geq y_G(u)$.

**Definition 6** Let $\tilde{T}_C \subseteq \tilde{T}_D$. Choose two Ω-soft sets $\tilde{T}_f$, $\tilde{T}_g$. Then $\tilde{T}_f = \tilde{T}_g$ if C=D, $\tilde{T}_C = \tilde{T}_D$ and $f_C = g_D$.

Let Pythagorean fuzzy numbers (PFNs) are denoted by $R = (d_R, y_R)$ [31]. Choose three PFNs $\theta = (d, y)$, $\theta_1 = (d_1, y_1)$, $\theta_2 = (d_2, y_2)$. We can give some basic operations as follows [30, 31]: For $\alpha > 0$,

$$\bar{\theta} = (y, d),$$

$$\theta_1 \oplus \theta_2 = (\sqrt{d_1^2 + d_2^2 - d_1^2 \cdot d_2^2},\ y_1 y_2),$$

$$\theta_1 \otimes \theta_2 = (d_1 d_2,\ \sqrt{y_1^2 + y_2^2 - y_1^2 \cdot y_2^2}),$$

$$\theta_1 \wedge \theta_2 = (\min\{d_1, d_2\}, \max\{y_1, y_2\}),$$

$$\theta_1 \vee \theta_2 = (\max\{d_1, d_2\}, \min\{y_1, y_2\}),$$

$$\alpha\theta = (\sqrt{1 - (1 - d^2)^\alpha},\ n^\alpha),$$

$$\theta^\alpha = (d^\alpha, \sqrt{1 - (1 - y^2)^\alpha}).$$

**Theorem 1** [36] Let $R = (d_R, y_R)$, $T = (d_T, y_T) \in K$ be two PFVs. Then, for $\alpha, \alpha_1, \alpha_2 > 0$,

  i.   $R \oplus T = T \oplus R$,
  ii.  $R \otimes T = T \otimes R$,
  iii. $\alpha(R + T) = \alpha R \oplus \alpha T$,
  iv.  $\alpha_1 R + \alpha_2 R = (\alpha_1 + \alpha_2)R$,
  v.   $(R \otimes T)^\alpha = R^\alpha \otimes T^\alpha$,
  vi.  $R^{\alpha_1} \otimes R^{\alpha_2} = R^{(\alpha_1 + \alpha_2)}$.

IFS, offered by Atanassov [2] is an extension of FS Theory [35]. IFS is characterized by a membership degree (MD) and a non-membership degree (ND) and therefore can indicate the fuzzy character of data in more detail comprehensively. The prominent characteristic of IFS is that it assigns to each element an MD and ND with their sum equal to or less than 1. However, in some practical DM processes, the sum of the MD and the ND to which an alternative satisfying a criterion provided by a decision-maker may be bigger than 1, but their square sum is equal to or less than 1.

Table 2 explains the difference between PFSs and IFSs.

**Table 2.** PFSs and IFSs

| IFSs | PFSs |
| --- | --- |
| $d + y \leq 1$ | $d + y \leq 1$ or $d + y \geq 1$ |
| $0 \leq d + y \leq 1$ | $0 \leq d^2 + y^2 \leq 1$ |
| $b = 1 - (d+y)$ | $b = \sqrt{1 - [d^2 + y^2]}$ |
| $b + y + b = 1$ | $b^2 + d^2 + y^2 = 1$ |

Therefore, Yager [32] proposed PFS characterized by an MD and an ND, which satisfies the condition that the square sum of its MD and ND is less than or equal to 1. Yager [33] gave an example to state this situation: A decision-maker gives his support for membership of an alternative is $\frac{\sqrt{3}}{2}$ and his against membership is $\frac{1}{2}$. Owing to the sum of two values is bigger than 1, they are not available for IFS, but they are available for PFS since $\left(\frac{\sqrt{3}}{2}\right)^2 + \left(\frac{1}{2}\right)^2 \leq 1$. Obviously, PFS is more capable than IFS to model the vagueness in the practical multi-criteria decision-making problems.

The main difference between PFNs and IFNS is their corresponding constraint conditions. Here, we observe that intuitionistic membership grades are all points under the line $d + y \leq 1$ and the Pythagorean membership grades are all points with $d^2 + y^2 \leq 1$.

One important implication of this is that it allows the use of the PFSs in situations in which we cannot use IFSs. An example of this would be a case in which a user indicates that their support for membership $x$ is $\frac{\sqrt{3}}{2}$ and their support against membership is $\frac{1}{2}$. As we noted these values are not allowable for intuitionistic membership grades but allowable as Pythagorean membership grades. Thus in this case, rather than requiring the user to change their information to satisfy the constraints of the IFS, we can use a PFS.

**PFS Method**

For PFNs, the mapping $SF: K \to [-1,1]$ is called score function, if

$$SF_R = d_R^2 - y_R^2 \qquad (1)$$

for all $R = (d_R, y_R) \in K$ [1,36].
For any two PFNs $R, T$,

$$\begin{cases} R \prec T, & \text{if } SF(R) < SF(T), \\ R \succ T, & \text{if } SF(R) > SF(T), \\ R \sim T, & \text{if } SF(R) = SF(T). \end{cases} \qquad (2)$$

As can be seen from (1), the score the larger the score $S_R$, the greater the PFN $R$ [8]. It should be noted that SF cannot differentiate some evidently distinct PFNs that have the same score. To explain this situation, we can give examples using (2): Take two PFNs $R, T$ as R = (0.481, 0.402) and T = (0.527, 0.456). Then $SF_R = 0.0687$ and $SF_T = 0.0697$. Again, for $R = (0.123, 0.123)$ and $T = (0.456, 0.456)$, we can write $R \sim T$. Therefore, if only the scoring function is used for comparison, it is not possible to make a comparison between these numbers. To overcome this problem, we can define a new function [1], as follows:

The mapping $AF: K \to [0,1]$ is called accuracy function, if

$$AF_R = d_R^2 + y_R^2 \qquad (3)$$

for all $R = (d_R, y_R) \in K$ [25].

Using the (1) and (3), for comparing PFVs, the following method is presented by Agarwal et al. [1]. For any two PFNs $R, T$, if $SF(R)=SF(T)$, then

$$\begin{cases} R \succ T, & \text{if } AF(R) > AF(T), \\ R \prec T, & \text{if } AF(R) < AF(T), \\ R \sim T, & \text{if } AF(R) = AF(T). \end{cases} \qquad (4)$$

For a binary relation $\leq_{(SF,AF)} \in K$ and $R, T \in K$, it can be written as

$$R \leq_{(SF,AF)} T \Leftrightarrow (SF_R < SF_T) \vee (SF_R = SF_T \wedge AF_R \leq AF_T).$$

Now, we can give the new definition:

**Definition 7** [15] The mapping $ES: K \to [0,1]$ is called expectation score function such that for all $R = (d_R, y_R) \in K$

$$ES_r = \frac{d_R^2 - y_R^2 + 1}{2}. \qquad (5)$$

**Definition 8** [15] For two PFNs $R, T \in K$ and the relation $\leq_{(ES,d)}$ on $K$, we have

$$R \leq_{(d,ES)} T \Leftrightarrow (d_R < d_T) \vee (d_R = d_T \wedge ES_R \leq ES_T).$$

By replacing the approval rates in Definition 8, the other relation $\leq_{(ES,d)}$ is written as follows:

**Definition 9** [15] For two PFNs $R, T \in K$ and the relation $\leq_{(ES,d)}$ on $K$, we have

$$R \leq_{(ES,d)} T \Leftrightarrow (ES_R < ES_T) \vee (ES_R = ES_T \wedge d_R \leq d_T).$$

The relationship with each other of $\leq(SF;AF)$ and $\leq_{(ES,d)}$ can be given as follows:

**Proposition 1** [15] Let $R$ and $T$ be PFNs in $K$. Then $R \leq_{(SF,AF)} T \Leftrightarrow R \leq_{(ES,d)} T$.

**Proposition 2** [15] For two PFNs $R, T, V \in K$ and $\alpha, \alpha_1, \alpha_2 > 0$. Then,

$$R \leq_{(ES,d)} V \Leftrightarrow T \oplus R \leq_{(d,ES)} T \oplus V.$$

**Algortihm**

Pythagorean fuzzy weighted averaging (PFWA) operator was given by Yager [34]. The following definition is based on the definition of PFWA of Yager.

**Definition 11** Let $\pi_i = (d_1, y_i)$ be PFVs in $K$, $K$, $\omega = \{\omega_1, \omega_2, \ldots, \omega_k\}^T$ is the weighted vector such that for $i =1,2,\ldots,k$, $\omega_i \in [0,1]$ with $\sum_{i=1}^{k} \omega_i = 1$. Then the mapping $\Omega_{PFWA}: K^n \to K$ given by

$$\Omega_{PFWA}(\pi_i) = \left( \sum_{i=1}^{k} \omega_i d_i, \sum_{i=1}^{k} \omega_i y_i \right)$$

is called PFWA operator.

**Definition 12** [33, 34] The mapping $PF: K^n \to K$ given by

$$PF_\omega(\pi_i) = \omega_1 \pi_1 \oplus \omega_2 \pi_2 \oplus \ldots \oplus \omega_k \pi_k = \left( \sqrt{1 - \prod_{i=1}^{k}(1-d_i^2)^{\omega_i}}, \prod_{i=1}^{k} n_i^{\omega_i} \right)$$

is called the $k$ dimensional PFWA operator.

Definition 12 can be used to simplify the computation concerning PFWA operators.

**Definition 13** [15] Let $F_P \in \Omega(E)$. Then,

$$AP_{F_P} = \oplus_{p \in P} \frac{ES(s_i)}{\sum_{p \in P} ES(s_i)} F_P$$

is called the aggregated Pythagorean fuzzy decision value(APFDV), where

$$ES(s_i) = \frac{\sum_{p \in P}[d(s_i)]^2 - \sum_{p \in P}[y(s_i)]^2 + 1}{2}.$$

**Algorithm:**

**Step 1:** The set of expert's P and the set S are recorded in the PFS table.
**Step 2:** The expectation values $ES(s_i)$ are calculated.

**Step 3:** The weights are found by

$$\omega = \frac{ES(s_i)}{\sum ES(s_i)}.$$

**Step 4:** For $k = 1, 2, \ldots, i$, the APFDVs are computed by

$$AP_{F_P}(p_k) = \bigoplus_{l=1}^{j} \frac{ES(s_i)(x_l)}{\sum_{p \in P} ES(s_i)(x_l)} F_{P(x_l)}(p_k).$$

**Step 5:** Rank $AP_{FP}(p_k)$, $(k=1,2,\ldots,i)$ descending under the order $\leq_{(d,ES)}$.

**Step 6:** Rank $(p_j)$, $(j=1,2,\ldots,k)$ correspondingly and output $p_k$ as the optimal decision, if $AP_{FP}(p_i)$ is largest PFV under the order $\leq_{(d,ES)}$.

**Application: The Effect of Music on Cognitive Development**

The most critical information for early childhood educators are developmental areas and developmental stages followed by the child according to age. Cognitive development plays a central role as an area that directly affects and directs other areas of development. Early childhood, when most of the brain development is completed, is a period in which the foundation of cognitive skills is laid. From this point of view, researches on the brain and cognitive field from the womb reveal that music has an important place in this process.

The results of the studies, which were constructed by listening to classical music in the womb or by singing a lullaby after birth, show the importance of music in terms of the development of the individual. In this context, some early childhood education approaches have identified one of their basic principles as language and rhythm. In addition, music activities play an important role in almost all early childhood education programs. From this point of view, it can be said that there is a sub-field called music education in early childhood.

Now, we will investigate the effect of music on the cognitive development of early childhood education. In Example 1, four experts related to early childhood education were chosen and four opinions about the cognitive development of children were given. The experts are studying research on the effect of music on the cognitive development of the early childhood period.

The ranking method proposed with the given algorithm will be used for the effects of music on children's cognitive development. An assessment will be made with the opinions of the four experts. As decision-makers, these experts will make this assessment according to the criteria in Example 1. Experts' opinions regarding these criteria will be listed by the solving procedure. This ranking will show the importance given by experts to these criteria.

We consider the values of Table 1. We compute the expectation values ES that reveal the weight vector (Table 3)

**Table 3.**

|  | s1 | s2 | s3 | s4 |
|---|---|---|---|---|
| $ES(s_i)$ | 0.89 | 0.47 | 0.68 | 0.72 |
| ω | 0.32246377 | 0.17028986 | 0.24637681 | 0.26086956 |

$\omega = \{0.32246377, 0.17028986, 0.24637681, 0.26086956\}^T$

to be used for calculating the APFDVs. The $AP_{FP}(p_k)$ is found as

$$AP_{FP}(p_i) = PF_\omega(\pi_i)(F_P(s_1)(p_i), F_P(s_2)(p_i), F_P(s_3)(p_i), F_P(s_4)(p_i)).$$

For example, $AP_{FP}(p_1) = (0.668, 0.501)$. From Table 4, we have,

$$AP_{FP}(p_3) \leq_{(d,ES)} AP_{FP}(p_1) \leq_{(d,ES)} AP_{FP}(p_4) \leq_{(d,ES)} AP_{FP}(p_2).$$

According to these results, the experts' opinion will be sorted as: $p_3 > p_1 > p_4 > p_2$.

**Table 4.** Measures

|  | APFDVs | $ES(AP_{FP}(p_i))$ | $SF(AP_{FP}(p_i))$ | $AF(AP_{FP}(p_i))$ |
|---|---|---|---|---|
| p1 | (0.668, 0.501) | 0.598 | 0.195 | 0.7 |
| p2 | (0.602, 0.641) | 0.476 | -0.048 | 0.77 |
| p3 | (0.800, 0.460) | 0.7142 | 0.43 | 0.8516 |
| p4 | (0.660, 0.494) | 0.595 | 0.2 | 0.68 |

**RESULTS AND DISCUSSION**

Schellenberg and Weiss [27] stated that there is a strong relationship between music tendency and general cognitive abilities, especially in childhood. He also emphasized that cognitive performance may increase as a result of the improvement of the general mood with the effect of music. Schellenberg, Nakata, Hunter, and Tamoto, [26] showed that the cognitive performance of 5-year-old children with different music genres can be increased. In the longitudinal study by Costa-Giomi [6], it was revealed that children who started music education before the age of 5 got significantly higher scores in spatial skills than children who started later or did not receive an education. Ho, Cheung, and Chan [9] found that verbal memory significantly differs according to the control group in boys aged 6-15 who receive music education. Ho, Cheung, and Chan [9] concluded that music training systematically influenced memory processing according to possible neuroanatomical changes in the left temporal lobe in their study of brain waves.

We applied, the entries in the PFS table, which is arranged according to the opinions of the experts, in the algorithm we obtained. Depending on the PFNs, the new method we proposed provides solutions to the decision analysis problems by the ranking of the PFNs. The results of the algorithm supported the data of the experts on the development of spatial-temporal skills of music education given in early childhood.

## CONCLUSION

In this study, a new decision-making algorithm and method were given. We used PFS in the given method. PFS was preferred because it is known that PFS gives clearer results than IFS. The effect of music on cognitive development in early childhood was examined with this decision-making method. In practice, the opinions of the experts about cognitive development and the results of the method we proposed were compared.

In this study, the expectation score function was used. Weights and thus APFDV are calculated with the values obtained from this function. The ranking is done with APFDV. Here, the values obtained from expert opinions are determined as follows: Whichever expert has given more opinions about the criteria, he/she has been in the ranking before. Again, whichever specialist has given fewer opinions remains behind the rankings. This is very suitable for real-life events. The rankings obtained from the algorithm of the study were the same as the rankings of the opinions of the experts.

## AUTHORSHIP CONTRIBUTIONS

Authors equally contributed to this work.

## DATA AVAILABILITY STATEMENT

The authors confirm that the data that supports the findings of this study are available within the article. Raw data that support the finding of this study are available from the corresponding author, upon reasonable request.

## CONFLICT OF INTEREST

The author declared no potential conflicts of interest with respect to the research, authorship, and/or publication of this article.

## ETHICS

There are no ethical issues with the publication of this manuscript.

## REFERENCES

[1] Agarwal M, Biswas KK, Hanmandlu M. Generalized intuitionistic fuzzy sets with applications in decision making. Appl Soft Compt 2013;13:3552–3566. [CrossRef]

[2] Atanassov K. Intuitionistic fuzzy sets. Fuzzy Sets Syst 1986;20:87–96. [CrossRef]

[3] Bardak, M. Game-based Learning. In: Gurol A (editors). Learning Approaches in Early Childhood. Istanbul: Efe Akademi Publications; 2018. p. 207–230. [Turkish]

[4] Bilhartz TD, Bruhn RA, Olson JE. The effect of early music training on child cognitive development. J Appl Develop Psychol 2000;20:615–636. [CrossRef]

[5] Cooper PK. It's all in your head: A meta-analysis on the effects of music training on cognitive measures in schoolchildren. Int J Music Educ 2020;38:321–336. [CrossRef]

[6] Costa-Giomi, E. The relationship between absolute pitch and spatial abilities. In: Woods C, Luck G, Brochard R, Seddon F, Sloboda JA (editors). Proceedings of the Sixth International Conference on Music Perception and Cognition. Keele, UK: Keele University, Department of Psychology; 2000.

[7] Feng F, Fujita H, Ali MI, Yager RR, Liu X. Another view on generalized ıntuitionistic fuzzy soft sets and related multiattribute decision making methods. IEEE Trans Fuzzy Syst 2018;27:474–488. [CrossRef]

[8] Greata J. An introduction to in early childhood education. New York: Thomson Delmar Learning; 2006.

[9] Ho Y, Cheung M, Chan AS. Music training improves verbal but not visual memory: Crosssectional and longitudinal explorations in children. Neuropsychology 2003;17:439–450. [CrossRef]

[10] Husain I, Rhemm A, Ahmad M. Analysis of the maximum age group of women affected by the divorce problem using fuzzy matrix method Int J Appl Math Stat 2018;57:36–46.

[11] Husain I, Aleen A. Fuzzy matrix approach to study the maximum age group of stressed students studying in higher education. Int J Emerg Technol 2021;12:31–35.

[12] Katarzyna B, Brenda BS. The effects of music instruction on cognitive development and reading skills-an overview. Bullet Council Res Music Educ 2011;189:89–104. [CrossRef]

[13] Kirişci M. Medical decision making with respect to the fuzzy soft sets. J Interdiscip Math 2020;23:767–776. [CrossRef]

[14] Kirişci M. A Case Study for medical decision making with the fuzzy soft sets. Afr Mat 2020;31:557–564. [CrossRef]

[15] Kirişci M. -soft sets and medical decision-making application. Int J Comput Math 2020;98:690–704. [CrossRef]

[16] Maji PK, Bismas R, Roy AR. Fuzzy soft set. J Fuzzy Math 2001;9:589–602.

[17] Maji PK, Bismas R, Roy AR. Intuitionistic fuzzy soft sets. J Fuzzy Math 2001;9:677–692.

[18] Maji PK, Biswas R, Roy AR. Soft set theory. Comput Math. Appl 2003;45:555–562. [CrossRef]

[19] Majumdar P, Samanta S. Similarity measure of soft sets. New Math Natural Comput 2008;4:1–12. [CrossRef]

[20] Molodtsov D. Soft set theory-first results. Comput Math. Appl 1999;37:19–31. [CrossRef]

[21] Natalia, S, Fernando, G, Cecilia, A. The Contribution of dynamic assessment to promote inclusive education and cognitive development of socioeconomically deprived children with learning disabilities. Transylvanian J Psychol 2013(Suppl):207–222.

[22] Nogaj AA. Emotional intelligence and strategies for coping with stress among music school students in the context of visual art and general education students. J Res Music Educ 2020;68:78–96. [CrossRef]

[23] Oakley L. Cognitive development. New York, USA: Routledge Press; 2004. [CrossRef]

[24] Peng X, Yang Y, Song J, Jiang Y. Pythagorean fuzzy soft set and its application. Comput Eng 2015;41:224–229.

[25] Peng X, Yang Y. Some results for Pythagorean fuzzy sets. Int J Intell Syst 2015;30:1133–1160. [CrossRef]

[26] Schellenberg EG, Nakata T, Hunter PG, Tamoto S. Exposure to music and cognitive performance: Tests of children and adults. Psychol Music 2007;35:5-19. [CrossRef]

[27] Schellenberg EG, Weiss MW. Music and cognitive abilities, In Deutsch D (editor). The Psychology of Music. Netherlands: Elsevier Academic Press; 2013. p. 499–550. [CrossRef]

[28] Shahzadi G, Akram A. Hypergraphs based on pythagorean fuzzy soft model. Math Comput Appl 2019;24:100. [CrossRef]

[29] Solso RL. Cognitive Psychology. Needham Heights: Allyn & Bacon; 1995.

[30] Topaç N. Investigating opinions of pre-school teachers and the parents who have pre-school children about the pre-school music education (master thesis). Istanbul: Marmara University; 2008. [Turkish]

[31] Vygotsky LS. The problem of the cultural development of the child. In: Van der Veer R, Valsiner J (editors). The Vygotsky Reader. Oxford: Basil Blackwell Ltd; 1994.

[32] Yager RR. Pythagorean fuzzy subsets, In: Proc Joint IFSA World Congress and NAFIPS Annual Meeting, Edmonton, Canada 2013:57–61. [CrossRef]

[33] Yager RR, Abbasov AM. Pythagorean membership grades, complex numbers, and decision making. Int J Intell Syst 2013;28:436–452. [CrossRef]

[34] Yager RR. Multi-criteria decision making with ordinal linguistic intuitionistic fuzzy sets for mobile apps. IEEE Trans Fuzzy Syst 2016;24:590–599. [CrossRef]

[35] Zadeh LA. Fuzzy sets. Inf Comp 1965;8:338–353. [CrossRef]

[36] Zhang XL, Xu ZS, Extension of TOPSIS to multi-criteria decision making with Pythagorean fuzzy sets. Int J Intell Syst 2014;29:1061–1078. [CrossRef]